\documentclass[useAMS]{mn2e}
\usepackage{times,graphicx,epsf}
\def\et{{\rm et al}~}
\title[MONTE CARLO SIMULATIONS OF STAR CLUSTERS III]
{Monte Carlo simulations of star clusters -- III. A million body
star cluster.}

\author[Mirek Giersz]{Mirek Giersz$\rm ^{1}$\thanks{E--mail: {\tt mig@camk.edu.pl}} \\
$\rm ^1$ Nicolaus Copernicus Astronomical Center, Polish Academy of Sciences, ul.
Bartycka 18, 00--716 Warsaw, Poland\\}

\begin{document}

\maketitle

\begin{abstract}
A revision of Stod\'o{\l}kiewicz's Monte Carlo code is used to
simulate the evolution of million body star clusters. The new method
treats each {\it superstar} as a single star and follows the
evolution and motion of all individual stellar objects. A survey of
the evolution of $N$--body systems influenced by the tidal field of
a parent galaxy and by stellar evolution is presented. The process
of energy generation is realized by means of appropriately modified
versions of Spitzer's and Mikkola's formulae for the interaction
cross section between binaries and field stars and binaries
themselves. The results presented are in good agreement with
theoretical expectations and the results of other methods. During
the evolution, the initial mass function (IMF) changes
significantly. The local mass function (LMF) around the half--mass
radius closely resembles the actual global mass function (GMF). At
the late stages of evolution the mass of the evolved stars inside
the core can be as high as $97\%$ of the total mass in this region.
For the whole system, the evolved stars can compose up to $67\%$ of
the total mass. The evolution of cluster anisotropy strongly depends
on initial cluster concentration, IMF and the strength of the tidal
field. The results presented are the first step in the direction of
simulating the evolution of real globular clusters by means of the
Monte Carlo method.
\end{abstract}

\begin{keywords}
gravitation -- methods: numerical -- celestial mechanics,
           stellar dynamics -- globular clusters: general
\end{keywords}

\section{INTRODUCTION.}

Dynamical modeling of real, large stelar systems, like globular
clusters or galactic nuclei, and understanding their evolution still
is a great challenge both for the theory and hardware/software.
Basically, there are two approaches. Direct $N$--body method, which
requires extremely large hardware requirements and very
sophisticated software, and statistical modeling based on the
Fokker--Planck and other approximations, which suffers from the poor
understanding of the validity of assumptions.

On the side of $N$--body simulations recent years brought a
significant progress in both hardware and software. First, parallel
(even massively parallel) computing has opened a route to gain
performance at relatively low cost and little technological
advancement. Secondly, the already successful GRAPE special purpose
computers (Sugimoto et al. 1990, Makino et al. 1997, Makino \& Taiji
1998, Makino 2002, Makino \& Hut 2003) have been developed in their
present generation (GRAPE--6) and aim at the 100 Tflops speed.
NBODY6++ (Spurzem 1999), the successor of Aarseth's NBODY6 code has
been ported on massively parallel, general purpose computers (CRAY
T3E and PC clusters) and was used for astrophysical problems related
to the dissolution of globular clusters (Baumgardt 2001, Baumgardt,
Hut \& Heggie. 2002) and the decay of massive black hole binaries in
galactic nuclei after a merger (Milosavljevic \& Merritt 2001,
Hemsendorf, Sigurdsson \& Spurzem 2002). Despite such progresses in
hardware and software there is still impossible to model directly
evolution of real globular cluster ($N \sim 10^6$) and galactic
nuclei ($N \sim 10^9$). Recent work by Baumgardt \et (2002) and
Baumgardt \& Makino (2002) has pushed the limits of present direct
modeling to about $10^5$ using both NBODY6++ and the GRAPE--6
special purpose hardware.

On the side of Fokker--Planck method with finite differences and an
anisotropic gaseous model, recent years brought great improvements.
Models can be used now to simulate more realistic stellar systems.
They can tackle: anisotropy, rotation, a tidal boundary, tidal
shocking by galactic disk and bulge, mass spectrum, stellar
evolution and dynamical and primordial binaries (Louis \& Spurzem
1991, Giersz \& Spurzem 1994, Takahashi 1995, 1996, 1997, Spurzem
1996, Takahashi \& Portegies Zwart 1998, 2000 hereafter TPZ, Drukier
\et 1999, Einsel \& Spurzem 1999, Takahashi \& Lee 2000, Kim et al.
2002, Kim, Lee \& Spurzem 2004, Fiestas, Spurzem \& Kim 2005).
Unfortunately, the Fokker--Planck approach suffers, among other
things, from the uncertainty of differential cross--sections of many
processes which are important during cluster evolution. It can not
supply detailed information about the formation and movement of all
objects present in clusters. Additionally, the anisotropic gaseous
model assumes a certain form of heat conductivity and closure
relations between the third order moments.

There is an elegant alternative to generate models of star clusters,
which can correctly reproduce the stochastic features of real star
clusters, but without really integrating all orbits directly as in
an $N$--body simulation. They rely on the Fokker--Planck
approximation and (hitherto) spherical symmetry, but their data
structure is very similar to an $N$--body model. These so--called
Monte Carlo models were first presented by H\'enon (1971, 1975),
Spitzer (1975) and later improved by Stod\'o{\l}kiewicz (1982, 1985,
1986) and in further work by Giersz (1997, 1998, 2001), and recently
by Rasio and his collaborators (Joshi, Rasio \& Portegies Zwart
2000, Watters, Joshi \& Rasio 2000, Joshi, Nave \& Rasio 2001,
Fregeau et al. 2003, G\"urkan, Freitag \& Rasio 2004), and Freitag
(Freitag 2000, Freitag \& Benz 2001, 2002). The basic idea is to
have pseudo--particles, which orbital parameters are given in a
smooth, self--consistent potential. However, their orbital motion is
not explicitly followed; to model interactions with other particles
like two--body relaxation by distant encounters or strong
interactions between binaries and other objects, a position of the
particle in its orbit, and further free parameters of the individual
encounter, are picked from an appropriate distribution by using
random numbers. The Monte Carlo scheme takes full advantage of the
established physical knowledge about the secular evolution of
(spherical) star clusters as inferred from continuum model
simulations. Additionally, it describes in a proper way the
graininess of the gravitational field and the stochasticity of real
$N$--body systems and provides, in a manner as detailed as in direct
$N$--body simulations, information about the movement of any objects
in the system. This does not include any additional physical
approximations or assumptions which are common in Fokker--Planck and
gas models (e.g. for conductivity). Because of this, the Monte Carlo
scheme can be regarded as a method which lies between direct
$N$--body and Fokker--Planck models and combines most of their
advantages. A hybrid variant of the Monte Carlo technique combined
with a gaseous model has been proposed by Spurzem \& Giersz (1996),
and applied to systems with a large number of primordial binaries by
Giersz \& Spurzem (2000) and  Giersz \& Spurzem (2003) and for
tidally limited systems by Spurzem \et (2005). The hybrid method
uses a Monte Carlo model for binaries or any other object for which
a statistical description, as used by the gaseous model, is not
appropriate due to small numbers of objects or unknown analytic
cross sections for interaction processes. The method is particularly
useful for investigating the evolution of large stellar systems with
a realistic fraction of primordial binaries, but could also be used
in the future to include, for example, the build up of massive stars
and blue stragglers by stellar collisions.

Very detailed observations of globular clusters which are available
at present or will become available in future (e.g. the Hubble Space
Telescope and the new very large terrestrial telescopes) have
extended and will extend our knowledge about their stellar content,
internal dynamics and the influence of the environment on cluster
evolution (Janes 1991, Djorgovski \& Meylan 1993, Smith \& Brodie
1993, Hut \& Makino 1996, Meylan \& Heggie 1997). This data covers
luminosity functions and derived mass functions, color--magnitude
diagrams, population and kinematical analysis, including binaries
and compact stellar evolution remnants, detailed two--dimensional
proper motion and radial velocity data, and tidal tails spanning
over arcs several degrees wide (Koch et al. 2004). A wealth of
information on "peculiar" objects in globular clusters (blue
stragglers, X--ray sources (high-- and low--luminosity), millisecond
pulsars, etc.) suggests a very close interplay between stellar
evolution, binary evolution and dynamical interactions. This
interplay is far from being understood. Moreover, recent
observations suggest that the primordial binary fraction in a
globular cluster can be as high as $15\%$ -- $38\%$ (Rubenstein \&
Bailyn 1997), and possible existence of intermediate--mass black
holes in the centers of some globular clusters (Gebhardt \et 2000).
With all these observational data such as King at al. (1998), Piotto
\& Zoccali (1999), Rubenstein \& Bailyn (1999), Ibata et al. (1999),
Piotto et al. (1999), Grillmair et al. (1999), Shara et al. (1998),
Odenkirchen et al. (2001), Hansen et al. (2002), Richer et al.
(2002) (to mention only a few papers), easily reproducible reliable
modelling becomes more important than before. Monte Carlo codes
provide all the necessary flexibility to disentangle the mutual
interactions between all physical processes which are important
during globular cluster evolution and to perform in a reasonable
time, detailed simulations of real globular clusters.

The ultimate aim of the project described here is to build a Monte
Carlo code (in the framework of the MODEST international
collaborations, {\bf http:/www.manybody.org/modest}), which will be
able to simulate the evolution of real globular clusters, as closely
as possible. In this paper (the third in the series) the Monte Carlo
code (which was discussed in detail in Papers I and II) is used to
simulate the evolution of stellar systems, which consist of
comparable number of stars and have mass comparable to real globular
clusters. The results of simulations will be compared with those of
Chernoff \& Weinberg (1990, hereafter CW), Vesperini \& Heggie
(1997, hereafter VH), Aarseth \& Heggie (1998, hereafter AH),
Baumgardt \& Makino (2002) and Lamers \et (2005).

The plan of the paper is as follows. In Section 2, a short
description of processes implemented into the Monte Carlo code will
be presented. In Section 3, the initial conditions will be discussed
and results of the simulation will be shown. And finally, in Section
4 the conclusions will be presented.

\section{MONTE CARLO METHOD.}

The Monte Carlo method can be regarded as a statistical way of
solving the Fokker--Planck equation. Its implementation presented in
Giersz (1998, hereafter Paper I) and Giersz (2001, hereafter Paper
II) is based on the orbit--averaged Monte Carlo method developed in
the early seventies by H\'enon (1971) and then substantially
improved by Stod\'o{\l}kiewicz (1986, and references therein), and
recently by Giersz (Paper I and II), by Freitag (Freitag \& Benz
2001, 2002) and by Rasio's group (Joshi \et 2000, 2001, Fregeau \et
2003, 2004, Fregeau, Chatterjee \& Rasio 2005, Freitag, G\"urkan \&
Rasio 2005, Freitag, Rasio \& Baumgardt 2005). The code is described
in detail in Paper I, which deals with simulations of isolated,
single--mass systems and in Paper II in which additional physical
processes were included: multi--mass systems and stellar evolution,
mass loss through the tidal boundary, formation of three--body
binaries and their subsequent interactions with field stars, and
interactions between binaries.

For the sake of completeness, a short description of the main
ingredients of the Monte Carlo model will follow. For more detailed
discussion of the code readers should refer to Paper I and II.

\subsection{Mass Spectrum and Stellar Evolution}

To facilitate comparison with results of other numerical simulations
of globular cluster evolution (e.g. CW, Fukushige \& Heggie 1995,
Giersz \& Heggie 1997, VH, AH, TPZ, JNR)  and for simplicity a
single power--low approximation of IMF was used.

\begin{equation}
N(m)dm = Cm^{-\alpha}dm,   \quad\quad\quad  m_{min} \leq m \leq
m_{max},
\end{equation}
where C and $\alpha$ are constants. The $m_{min}$, minimum stellar
mass and $m_{max}$, maximum stellar mass were chosen to $0.1
M_\odot$ and $15.0 M_\odot$, respectively. However, it should be
noted that observations give more and more evidence that the mass
function in globular/open clusters and for field stars as well is
not a simple power--law, but it could be rather approximated by a
composite power--law (e.g. Kroupa, Tout \& Gilmore 1993, Kroupa
2002). To describe the mass loss due to stellar evolution the same
simplified stellar evolution model as adopted by CW was used. More
sophisticated models of stellar and binary evolution based on
fitting analytical formulae to the evolution tracks for stars and
binaries were developed by Portegies Zwart \& Verbunt (1996),
Hurley, Pols \& Tout (2000), Hurley, Tout \& Pols (2003) and
Belczy\'nski, Kalogera \& Bulik (2002). These models, among others,
were used in population synthesis codes (e.g. Belczy\'nski \& Taam
2004), and simulations of the evolution of real star cluster M67
(e.g. Hurley {\it et al} 2005).

The initial masses of stars are generated from the continuous
distribution given in equation (1). This ensure a natural spread in
their lifetimes. To ensures that the cluster remains close to virial
equilibrium during rapid mass loss due to stellar evolution, special
care is taken, that no more than $3\%$ of the total cluster mass is
lost during one overall time--step.

\subsection{Tidal Stripping}

For tidally limited star cluster the rate of mass loss is much
higher than for isolated systems. This is connected with the tidal
stripping of mass across the system boundary by the gravitational
field of a parent galaxy. For isolated system mass loss is
attributed mainly to rare strong interactions in the dense, inner
part of the system. In the present Monte Carlo code a mixed
criterion is used to identify escapers: a combination of apocentre
and energy--based criteria (see Paper II). This means that no
potential escapers are kept in the system. Stars regarded as
escapers are instantaneously lost from the system. This is in
contrast to $N$--body simulations where stars need time proportional
to the dynamical time to be removed from the system. Recently,
Baumgardt (2000) showed that stars with energy greater than the
tidal cutoff energy $E_t = -GM/r_t$, in the course of escape, can
again become bound to the system, because of distant interactions
with field stars. A similar conclusion was reached by Fukushige \&
Heggie (2000), who showed that the escape time--scale can be very
long (much larger than the half--mass relaxation time).
Additionally, during the final stages of cluster evolution or for
clusters with initially low central concentration, the mass loss
across the tidal boundary can become unstable when too many stars
are removed from the system at the same time. To properly follow
these stages of evolution the time--step has to be decreased to
force smaller mass loss.

\subsection{Three--Body Binaries and their interactions}

In the present Monte Carlo code (as it was described in Paper I) all
stellar objects, including binaries, are treated as single {\it
superstars}. This allows to introduce into the code, in a simple and
accurate way, processes of stochastic formation of binaries and
their subsequent stochastic interactions with field stars and other
binaries. The procedure of binary formation in three--body
interactions is in great detail described in Paper II. It relies on
the observation that the probability that the masses of the three
stars involved in the interaction are $m_1$, $m_2$ and $m_3$ is
$n_1n_2n_3/n^3$ (where $n_1$, $n_2$, $n_3$ and $n$ are number
densities of three interacting stars and the total number density,
respectively) and the rate of binary formation is proportional to
$n_1n_2n_3$. These considerations lead to the formula for the
probability of binary formation which depends only on the local
total number density instead of local number densities of each mass
involved in the interaction (see Equation 7 in Paper II). This
procedure substantially reduces fluctuations in the binary formation
rate. The determination of the local number density in the Monte
Carlo code, particularly for multi--mass systems, is a very delicate
and difficult matter (Paper I).

The energy generation in binary -- field star interactions is
computed according to the (appropriately modified for multi--mass
case) semi--empirical formula of Spitzer (Spitzer 1987).
Unfortunately, for multi--component systems there is no simple
semi--empirical formula which can fit all numerical data (Heggie,
Hut \& McMillan 1996). The total probability of the interaction
between a binary of binding energy $E_b$ consisting of mass $m_1$
and $m_2$ and a field star of mass $m_3$ was deduced from Equations
(6-23), (6-11) and (6-12) presented in Spitzer (1987) and results of
Heggie (1975) with an appropriately adjusted coefficient (the total
probability computed is such a way gives correct value for the
equal--mass case).

\begin{equation}
P_{3bf} = {5\sqrt2\pi A_sG^2m_1^2m_2^2\sqrt
{m_{123}}n\over{8\sqrt3\sqrt{m_{12}}\sqrt{m_3} \sqrt{m_a} \sigma
E_b}}\Delta t,
\end{equation}
where $m_{12} = m_1 + m_2$, $m_{123} = m_{12} + m_3$, $m_a$ is the
mean stellar mass in a zone, and $\sigma$ is the one dimensional
velocity dispersion. This procedure is, of course, oversimplified
and in some situations can not give correct results, for example,
when a field star is very light compared to the mass of the binary
components.

The implementation of interactions between binaries in the code is
based on the method described by Stod\'o{\l}kiewicz (1986) and then
improved in Paper II. Only strong interactions are considered, and
only two types of outcomes are permitted: heavier binary and two
single stars, or two binaries in a hyperbolic relative orbit. Stable
three--body configurations are not allowed. In the case when
binaries are only formed in dynamical processes, and only a few
binaries are present, at any time, in the system, it is very
difficult to use the binary density (Giersz \& Spurzem 2000) to
calculate the probability of a binary--binary interaction. Given
binary can hit only binaries (regarded as targets) whose actual
distance from the cluster center lies between its pericenter and
apocenter. This leads to the following formula for the
binary--binary interaction probability.

\begin{equation}
P_{3b3b} = {w\over |v_r|} {p^2\over {2r^2T}} \Delta t,
\end{equation}
where $w$ is the relative velocity between interacting binaries,
$|v_r|$, $r$, $T$ and $p$ are the radial velocity, radial position,
orbital period of a binary in the system and impact parameter (equal
to $2.5$ times the semi--major axis of the softer binary),
respectively. The outcome of the interaction is as follows (see
details in Stod\'o{\l}kiewicz 1986, Paper II and Mikkola 1983,
1984):
\begin{itemize}
  \item Two binaries in a hyperbolic relative orbit ($12\%$ of
  all interactions). Energy generation is equal to $0.4$ times
  the binding energy of the softer binary.
  \item One binary and two escapers ($88\%$ of all interactions).
  The softer binary is destroyed and the harder binary increases its
  binding energy by the amount equal to $0.516$ times the sum of
  binding energies of binaries.
\end{itemize}

The procedures described above are very uncertain in regard to the
amount of energy generation. To cure this problem, it is planned in
the near future to introduce into the code numerical procedures
(based on Aarseth's NBODY6 code), which can numerically integrate
the motion of three-- and four--body subsystems. A similar procedure
was introduced, with success, into the Hybrid code (Giersz \&
Spurzem 2000, 2003).

\section{RESULTS}

In Paper II a survey of the evolution of $N$--body systems
influenced by the tidal field of a parent galaxy, by stellar
evolution and by energy generation in interactions between binaries
and field stars and binaries themselves was discussed. Here, the
first Monte Carlo simulations of million body systems are presented.
The results will be discussed from the point of view of system
parameters which can be verified by observational data. Among
others, evolution of LMF and GMF, anisotropy, total mass, mean mass,
mass segregation and density profiles will be presented.

\subsection{Initial Models}

The initial conditions were chosen in a similar way as in the
``collaborative experiment'' (Heggie \et 1999) and Paper II. The
positions and velocities of all stars were drawn from the King model
(King 1966). All models have the same total number of stars $N_T =
1,000,000$ and the same tidal radius $r_t = 33.57$ pc. Masses of
stars are drawn from the power--law mass function according to
Equation (1). The minimum mass was chosen to be $0.1 M_\odot$ and
the maximum mass $15 M_\odot$. Two different values of the
power--law index were chosen: $\alpha = 2.35$ and $3.5$. The sets of
initial King models were characterized by $W_0 = 3$, $5$ and $7$.
All models are listed in Table 1.

\begin{table}
\begin{center}
\caption{Models $^a$}
\begin{tabular}{|l|l|l|l|l|l|l|l|} \hline \hline
  Model & $W_0$ & $\alpha$ & $M_T$ & $R_G$ & $r_{t_M}$ & $t_{scale}$ & $F$\\ \hline
  W3235 & 3 & 2.35 & 319305 & 2.00 & 3.1311 & 79797 & 4.62 \\
  W335 & 3 & 3.5 & 166577 &  2.77 & 3.1311 & 61418 & 3.34\\
  W5235$^b$ & 5 & 2.35 & 319305 & 2.00 & 4.3576 & 48602 & 4.62 \\
  W7235$^b$ & 7 & 2.35 & 319305 & 2.00 & 6.9752 & 23999 & 4.62 \\ \hline
  \multicolumn{8}{l}{} \\
  \multicolumn{8}{l}{$^a$ $M_T$ is the total cluster mass in solar mass, $r_{t_M}$ is the tidal}\\
  \multicolumn{8}{l}{radius in Monte Carlo units, $t_{scale}$ is the time scaling factor to}\\
  \multicolumn{8}{l}{scale simulation time to physical time in $10^6$ yrs, $R_G$ is the}\\
  \multicolumn{8}{l}{distance from the Galactic Center in kpc, $F$ in $10^4$ (Chernoff}\\
  \multicolumn{8}{l}{\& Weinberg 1990). The first entry after $W$ describes the King model}\\
  \multicolumn{8}{l}{and the following numbers the mass function power--law index.}\\
  \multicolumn{8}{l}{$^b$ two models with different initial random seed were simulated.}\\
  \end{tabular}
\end{center}
\end{table}
$F$ is the parameter (introduced by CW), which is proportional to
the half--mass relaxation time.

\begin{equation}
F \equiv {M_T \over {M_{\odot}}} {R_G \over {kpc}} {220 km s^{-1}
\over {v_g}} {1 \over{lnN}},
\end{equation}
where $R_G$ is the distance of the globular cluster to the Galactic
Center, and $v_g$ is the circular speed of the cluster around the
Galaxy. In order to scale the Monte Carlo time to physical units the
following formula is used

\begin{equation}
{t_{scale}\over 10^6 yrs} = 14.91\left({r_t\over
r_{t_M}}\right)^{1.5} {N_T \over {\sqrt{M_T} ln (\gamma N_T)}},
\end{equation}
where $r_{t_M}$ is the tidal radius in Monte Carlo units, $M$ is the
total cluster mass and $r_t$ is the tidal radius, both are in solar
mass and pc, respectively and $\gamma$ is equal to $0.11$ as for
single--mass systems (Giersz \& Heggie 1994). However, for
multi--mass systems it could be much smaller (Giersz \& Heggie 1996,
1997). The initial model is not in virial equilibrium, because of
statistical noise and because of masses, which  are assigned
independently from the positions and velocities. Therefore, the
model has to be initially rescaled to fulfill the virial equilibrium
condition. During all the simulations the virial ratio is kept
within $< 2\%$ of the equilibrium value (for the worst case within
$<5\%$). Standard $N$--body units (Heggie \& Mathieu 1986), in which
the total mass $M = 1$, $G = 1$ and the initial total energy of the
cluster is equal to $-1/4$, have been adopted for all runs. Monte
Carlo time is equal to $N$--body time divided by $N_T/ln(\gamma
N_T)$. In the course of evolution, when the cluster looses mass, the
tidal radius changes according to $r_t \propto M^{1/3}$.

Finally, a few words about the efficiency of the Monte Carlo code
presented here. The simulation of a million body systems took about
two/three weeks (depending on an initial model) on a Pentium 4, 2
GHz PC. This is still much faster than the biggest direct $N$--body
simulations performed on GRAPE--4 (Teraflop special--purpose
hardware). Nevertheless, the speed of the code is not high enough to
perform large scale survey simulations in a reasonable time. It is
clear, that to simulate the evolution of real globular clusters with
all physical processes in operation, a substantial speed--up of the
code is needed. This can be done, either by parallelizing the code
(in a similar way to JNR), introducing a more efficient way of
determining the new positions of {\it superstars}, introducing {\it
superstars} which contain a different number of stars, or by using a
hybrid code (as was done by Giersz \& Spurzem 2000, 2003). However,
the Monte Carlo code has presently a great relative advantage over
the $N$--body code for simulations with a large number of primordial
binaries. Primordial binaries substantially downgrade the
performance of $N$--body codes on supercomputers or on
special--purpose hardware. They can be introduced into the Monte
Carlo code in a natural way, practically without a substantial loss
of performance (Giersz \& Spurzem 2000, 2003).

\subsection{Global evolution.}

The rate of cluster evolution and indirectly the strength of the
tidal field is described by the parameter $F$ (Equation. 4).
Generally, the greater the value of $F$, the longer the relaxation
time and the slower the cluster evolution. The same relation is true
also for the distance of a globular cluster to the Galactic Center,
$R_G$ (keeping the constant total cluster mass $M_T$ and number of
stars $N_T$). The greater the $F$, the larger the $R_G$. As one can
see from the Table 1 distance $R_G$ is rather small This means that
the galactic tidal field is very strong and globular cluster
evolution is very fast. These initial conditions were chosen mainly
to speed up the simulations and discussed models similar to models
considered by VH.

\begin{figure}
\epsfverbosetrue
\begin{center}
\leavevmode \epsfxsize=80mm \epsfysize=60mm \epsfbox{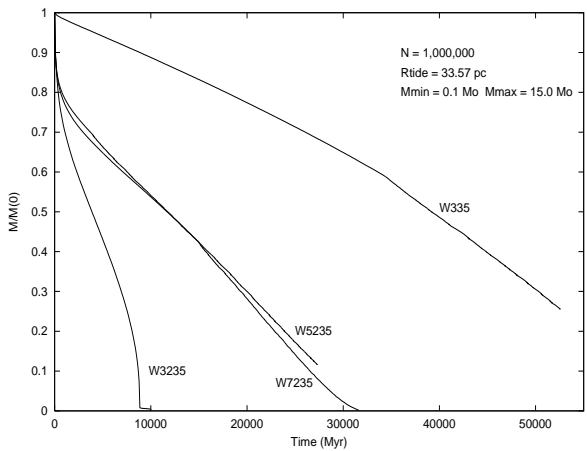}
\end{center}
\caption{Evolution of the total mass for models W7235, W5235, W3235
and W335.}
\end{figure}
Figure 1 presents the evolution of the total mass for all discussed
models. Three phases of evolution are clearly visible. The first
phase is connected with the violent mass loss due to stellar
evolution of the most massive stars. Then, there is the long phase
connected with gradual mass loss due to tidal stripping. And
finally, there is the phase connected with the tidal disruption of a
cluster. The last phase is well visible only in model W7235, which
evolution is followed long enough. In model W3235 on can see a phase
when the total cluster mass is nearly constant and very small (about
a few thousands of the initial mass). This is an artificial effect
connected with the fact that the Monte Carlo code can not properly
deal with systems with the total binding energy close to zero and
consist of only a few dozen particles. It is interesting to note,
that the second phase (gradual tidal stripping) can be divided into
two subphases, clearly separated by the core collapse time. For the
post collapse phase the rate of mass loss is slightly steeper than
in the collapse phase. This can be connected with the fact that hard
interactions, which involve binaries, start to operate introducing
an additional mechanism of the mass loss (escaping stars and
binaries). Models with the same mass--function ($\alpha \leq -2.35$)
and with different $W_0$ show very similar evolution during the
phase of tidal stripping (except W3235). It seems that the initial
mass loss due to stellar evolution is not sufficiently strong to
substantially change the initial cluster structure. Model W3235 is
not initially concentrated strongly enough to survive without
substantial structure changes, the violent mass loss event. Figure 1
presents qualitatively very similar features as figures shown by VH,
Paper II, Baumgardt \& Makino (2002) and Lamers \et (2005), Figure
1, Figures 1-3, Figure 1 and Figure 2, respectively.

\begin{figure}
\epsfverbosetrue
\begin{center}
\leavevmode \epsfxsize=80mm \epsfysize=60mm \epsfbox{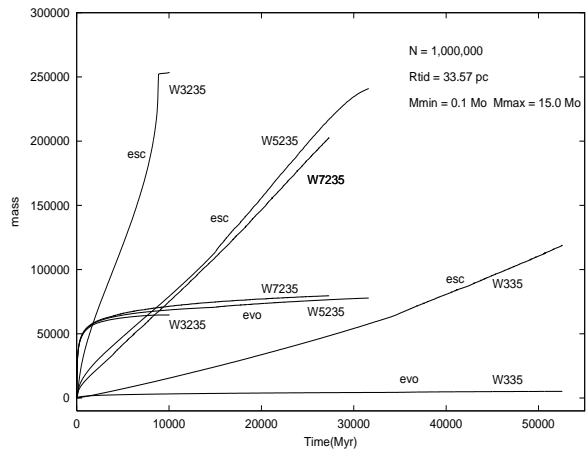}
\end{center}
\caption{Mass loss due to stellar evolution (evo) and tidal
stripping (esc) for models W7235, W5235, W3235 and W335}
\end{figure}
The total mass loss consists of mass loss due to stellar evolution
($evo$) and mass loss connected with tidal stripping ($esc$).
Results presented in Figure 2 are in a good qualitative agreement
with the $N$-body results of VH (Figure 2), taking into account that
$\alpha$ adopted by VH is $-2.5$ instead $-2.35$. According to
Equation 11 of VH, the evolution of the total mass of the cluster
can be fitted by a simple analytical expression, which consists of
two terms: $evo$ and $esc$. The amount of mass loss because of $esc$
can be fitted (as it was shown in VH) by a straight line with a
coefficient $\beta$. The values of $\beta$ listed by VH are 0.828
and 0.790 for models $W_0=7$, and $W_0=5$, respectively. The
equivalent values for the Monte Carlo models are 0.999 and 0.943.
The bigger values of $\beta$ for the Monte Carlo models can be
attributed mainly to the different: escape criterion, power--low
index of the initial mass function and slight increase of the rate
of mass loss due to binary activities (see Figure 1). As it was
stated in Section 2.2, the escape criterion adopted in this paper
(together with the instantaneous removal of escapers) leads to a
higher escape rate than in $N$--body simulations. Additionally, less
steep initial mass function causes stronger mass loss due to stellar
evolution and consequently leads to more escapers than the steeper
mass function adopted by VH. Mass loss connected with stellar
evolution dominates the initial phase of cluster evolution. Then the
rate of stellar evolution substantially slows down and escape due to
tidal stripping takes over. During this phase of evolution the rate
of mass loss due to stellar evolution is nearly constant, and higher
for shallower mass functions (as expected). Energy carried away by
stellar evolution events dominates the energy loss due to tidal
stripping, even though the tidal mass loss is higher.

It is worth to note, that all models except W3235 show clear
gravothermal oscillations in the post collapse phase. The
oscillations become less and less visible when the cluster is
approaching the final stages of dissolution.

Generally, agreement throughout the evolution between the results
presented here and those by AH, VH, Baumgardt \& Makino (2002) and
Lamers \et (2005) for $N$--body models, TPZ for 2--D Fokker--Planck
models and by JNR and Paper II for Monte Carlo models is rather
good. In all cases, the qualitative behaviour is very similar.
Nevertheless, the treatment of escapers for a tidally limited system
proposed by Spurzem \et (2005) can be implemented to properly ando
more accurately follow the mass loss due to tidal stripping and the
final stages of the cluster evolution, just before the dissolution.

At the end of this section the evolution of the density profiles for
model W7235 will be discussed (see Figure 3). The very large number
of particles in models allows substantial reduction in statistical
fluctuations and computation (with high precision) of the density
profiles in the course of cluster evolution. From the theoretical
considerations it follows that, during the collapse phase the nearly
isothermal core is building up. At the time of core bounce (about
3.5 Gyr) the density profile reaches the theoretically predicted
power--low index equal to $-2.2$. Then this index is preserved
during the further post--collapse evolution. For models W7235, W5235
and W335 the evolution of density profiles is practically the same.
Only for model W3235, which undergoes strong mass loss due to
stellar evolution and tidal stripping the density profiles do not
show any signs of power--low core development. The initial shape of
the profile is practically preserved (taking into account the
reduction of central density and tidal radius).

\begin{figure}
\epsfverbosetrue
\begin{center}
\leavevmode \epsfxsize=80mm \epsfysize=60mm \epsfbox{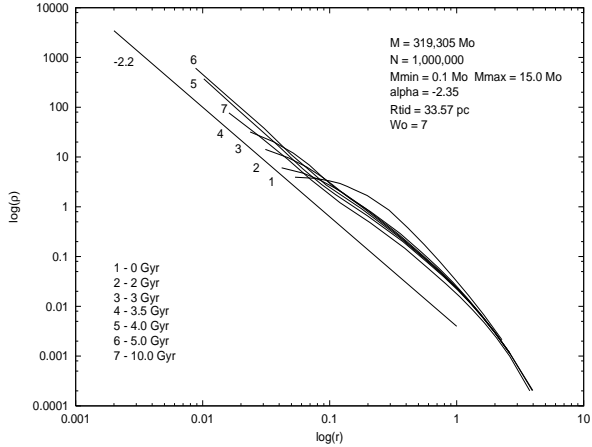}
\end{center}
\caption{Evolution of the density profiles for model W7235. The
times (in Gyr) at which density profiles are plotted are: 0, 2, 3,
3.5, 5, 6 and 7. The theoretically predicted density profile in the
core plotted as a straight line labeled by -2.2.}
\end{figure}

\subsection{Anisotropy evolution.}

The degree of velocity anisotropy is measured by a quantity

\begin{equation}
{\cal A} \equiv 2 - {2\sigma ^2_t \over {\sigma ^2_r}},
\end{equation}
where $\sigma_r$ and $\sigma_t$ are the radial and tangential
one--dimensional velocity dispersions, respectively. For isotropic
systems $\cal A$ is equal to zero. Systems preferentially populated
by radial orbits have $\cal A$ positive and systems preferentially
populated by tangential orbits have $\cal A$ negative. The evolution
of anisotropy is presented in Figures 4 to 6 for models W3235, W5235
and W7235, respectively. As can be seen in Figure 4, for systems
which can not survive the violent initial mass loss without strong
structural changes (see discussion in the previous section),
anisotropy stays close to zero with very large fluctuations. The
system does not live long enough and the tidal stripping is strong
enough to prevent the developing of substantial positive or negative
anisotropy. For model W5235 the cluster is not initially very
strongly concentrated, so the mass loss due to stellar evolution and
consequently by tidal stripping prevents the development of positive
anisotropy in the outer and middle parts of the system (see Figure
5). From the very beginning anisotropy for these regions becomes
negative. Stars, which are preferentially on radial orbits, escape
leaving mainly stars on tangential orbits. Finally, just before
cluster disruption, the anisotropy in the whole system becomes
slightly positive. At that time of cluster evolution most stars have
radial orbits. Clusters concentrated stronger (W7235), develop from
the very beginning a small positive anisotropy in the outer and
middle parts of the system (see Figure 6). In the course of
evolution the amount of anisotropy in the outer parts of the system
is reduced substantially by tidal stripping, as stars on radial
orbits escape preferentially. As tidal stripping exposes deeper and
deeper parts of the system, the anisotropy (for large Lagrangian
radii) gradually decreases and eventually becomes negative. At the
same time the anisotropy in the middle and inner parts of the system
stays close to zero.

It is interesting to note that for clusters which undergo core
collapse (W5235 and W7235) the anisotropy in the inner parts of the
system becomes slightly  positive just around the collapse time.
This behaviour can be attributed to the binary activities, which are
mainly concentrated in the cluster core. Interactions between
binaries and field stars (relaxation and energy generation) put
stars on more radial orbits.

\begin{figure}
\epsfverbosetrue
 \begin{center}
\leavevmode \epsfxsize=80mm \epsfysize=60mm \epsfbox{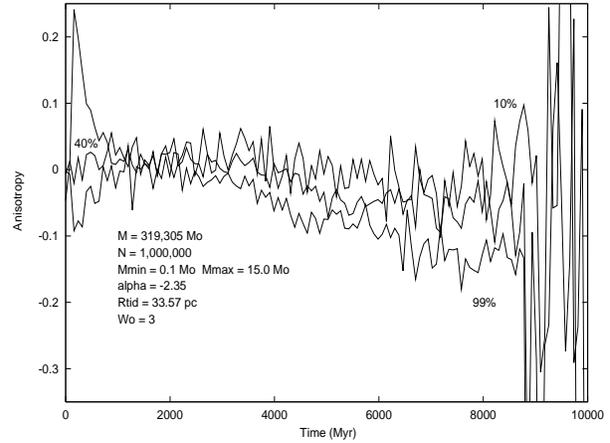}
\end{center}
\caption{Evolution of the anisotropy for 10\%, 40\% and 99\%
Lagrangian radii for model W3235.}
\end{figure}
\begin{figure}
\epsfverbosetrue
 \begin{center}
\leavevmode \epsfxsize=80mm \epsfysize=60mm \epsfbox{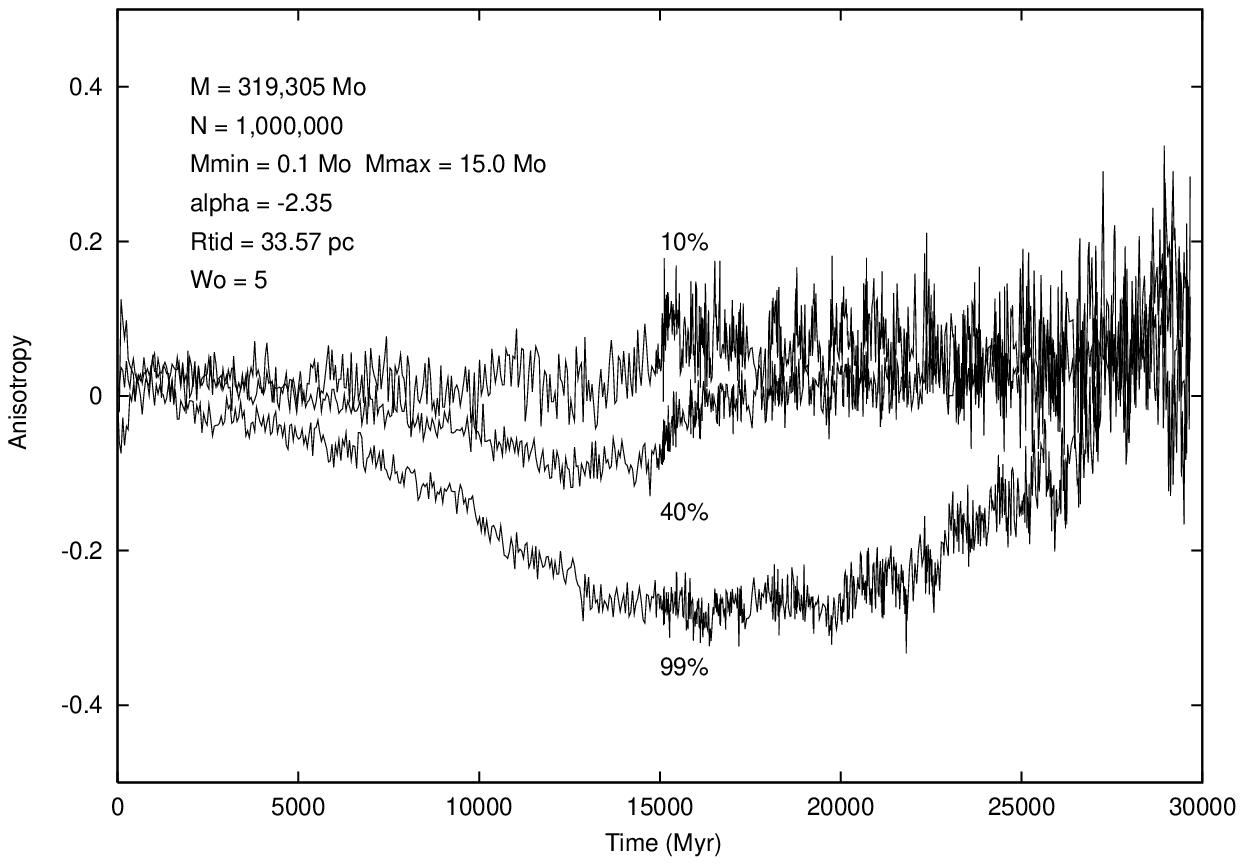}
\end{center}
\caption{Evolution of the anisotropy for 10\%, 40\% and 99\%
Lagrangian radii for model W5235.}
\end{figure}
\begin{figure}
\epsfverbosetrue
 \begin{center}
\leavevmode \epsfxsize=80mm \epsfysize=60mm \epsfbox{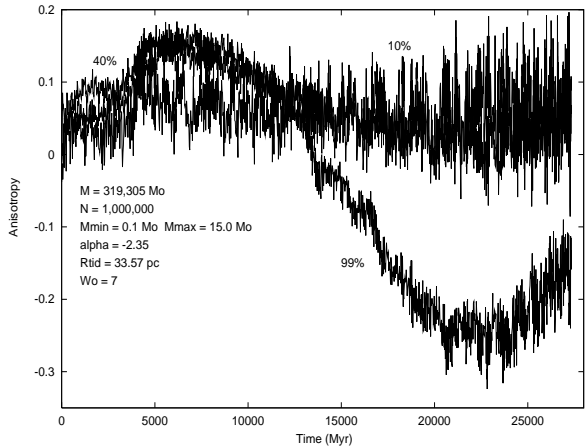}
\end{center}
\caption{Evolution of the anisotropy for 10\%, 40\% and 99\%
Lagrangian radii for model W7235.}
\end{figure}

The anisotropy of main--sequence stars shows very similar behaviour
to that discussed above. Mass segregation forces white dwarfs and
neutron stars (most massive stars) to preferentially occupy the
inner parts of the system. Therefore the anisotropy for them is much
more modest than for the main--sequence stars, and stays close to
zero. Anisotropy evolution agrees very well with the results of
Paper II and qualitatively with the results obtained by Takahashi
(1997) and Takahashi \& Lee (2000).

\subsection{Mass segregation.}

Three main effects may be expected to govern the evolution of the
mass function in models studied in this paper.
First, there is a period of violent mass loss due to sellar
evolution of the most massive stars. It takes place mainly during
the first few hundred million years and its amplitude strongly
depends on the slope of the mass function. Secondly, there is the
process of rapid mass segregation, caused by two--body distant
encounters (relaxation). It takes place mainly during
core collapse and is relatively
unaffected by the presence of a tidal field. Thirdly, there is the
effect of the tidal field itself, which becomes more important after
core collapse or even earlier (for models with a shallow mass
function and with low concentration). Because the tides
preferentially remove stars from the outer parts of the system,
which are mainly populated by low--mass stars (mass segregation), one
can expect that the mean mass should increase
as evolution proceeds, making allowance for stellar evolution.

The basic results are illustrated in Figure 7 for the total mean
mass inside the $10\%$ and $50\%$ Lagrangian radii and the tidal
radius, and in Figure 8 for the mean mass of main--sequence stars
and white dwarfs inside $10\%$ Lagrangian radius for model W5235.
The violent and strong mass loss due to stellar evolution is
characterized by an initial decrease of the mean mass. The evolution
of the most massive stars will remove a substantial amount of mass
from the system and, consequently, greatly lower the mean mass in
the whole system. This behaviour is visible also in the other
models, but with one exception. The mean mass inside the $10\%$
Lagrangian radius for model W7235 is increasing instead of
decreasing. For this model, the initial concentration is high enough
to force very quick mass segregation in the central part of the
system. The most massive main sequence stars (not evolved yet):
white dwarfs and neutron stars, sink into the center. For model
W3235 for most of the time the mean mass stays nearly constant for
the whole system. Just before the cluster dissolution it
substantially increases. For all models tidal stripping connected
mainly with removal of less massive stars forces the mean mass in
the middle and outer parts of the system to steadily increase. The
much faster increase of the mean mass inside the $10\%$ Lagrangian
radius (mass segregation) than in the outer Lagrangian radii ($50\%$
and $99\%$) during the collapse phase nearly stops at the core
bounce time. The subsequent evolution is characterized by the slower
rate of increase of the mean mass (close to the time of the core
bounce the mean mass is practically constant). This is in good
agreement with results obtained by Giersz \& Heggie (1997) for small
$N$--body simulations and for Monte Carlo simulations presented in
Paper II. The reason for the nearly constant mean mass is unclear.
The slow increase of the mean mass can be attributed to the binary
activity, which leads to removal of some low mass stars from the
core. For all models the effect of tides manifests itself by a
gradual increase of the mean mass in the halo.

\begin{figure}
 \epsfverbosetrue
  \begin{center}
 \leavevmode
\epsfxsize=80mm \epsfysize=60mm
 \epsfbox{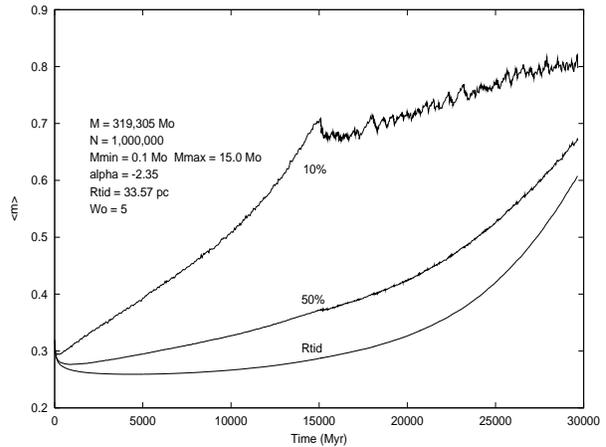}
\end{center}
\caption{Evolution of the mean mass inside 10\%, 40\% Lagrangian
radii and $r_{tid}$ for model W5235.}
\end{figure}
\begin{figure}
\epsfverbosetrue
 \begin{center}
\leavevmode \epsfxsize=80mm \epsfysize=60mm \epsfbox{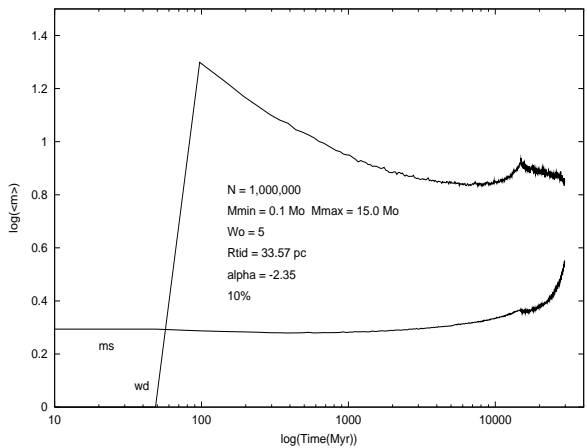}
\end{center}
\caption{Evolution of the mean mass for 10\% Lagrangian radius
for model W5235; $ms$ and $wd$ means mean--sequence stars and white
dwarfs, respectively.}
\end{figure}

The evolution of the mean mass for main--sequence stars and white
dwarfs for the inner $10\%$ Lagrangian radii is shown on Figure 8.
When less and less massive main--sequence stars finish their
evolution as less and less massive white dwarfs, the mean mass of
white dwarfs decreases with time. The newly created less massive
white dwarfs affect the mean mass of white dwarfs more strongly for
the steeper mass function than for the shallower one. For the
steeper mass function a smaller number of massive white dwarfs is
created comparable to the shallower mass function. At the time
around the core bounce the average mass of white dwarfs increases.
This is connected with the decreasing core size and continuing mass
segregation. Deeper in the cluster center a higher fraction of
massive white dwarfs is present. In the post collapse phase the
white dwarfs average mass is decreasing. Binaries start to be
created mainly from the most massive stars (neutron stars and white
dwarfs) deep in the core. They dynamically interact with other stars
and in the course of evolution are removed from the system together
with some massive white dwarfs. This leads to the decrease of the
mean mass of white dwarfs.

The mean mass of main--sequence stars initially slightly decreases,
which is connected with stellar evolution of the most massive stars.
Then, there is a period of the gradual increase of the mean mass due
to mass segregation. At late phases of cluster evolution the rate of
increase of the mean mass speeds up. Around the core bounce time one
can observe the break of the rate of increase of the main--sequence
average mass. This can be again attributed to binary activities.

The evolution of the total mean stellar mass in the different places
of the system (Figure 7) and the evolution of the average mass
inside $10\%$ Lagrangian radius for main--sequence stars and white
dwarfs (Figure 8) agree very well with the results obtained in Paper
II.

\begin{figure}
\epsfverbosetrue
 \begin{center}
\leavevmode \epsfxsize=80mm \epsfysize=60mm \epsfbox{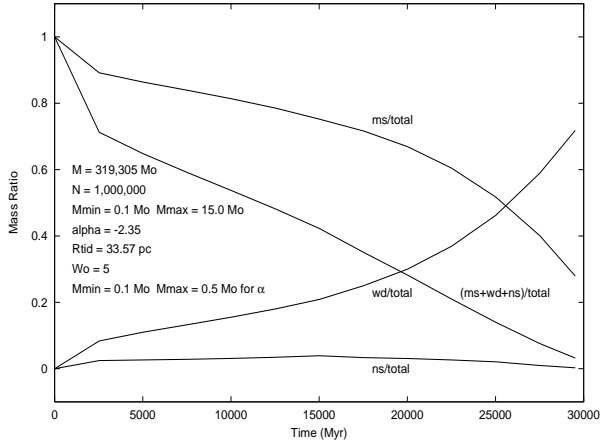}
\end{center}
\caption{Evolution of the mass ratio of main--sequence stars (ms),
white dwarfs (wd) and neutron stars/black holes (ns) to the actual
total cluster mass and the evolution of the ratio of the actual
total mass (ms+wd+ns) to the initial total mass for model W5235.}
\end{figure}
\begin{figure}
\epsfverbosetrue
 \begin{center}
\leavevmode \epsfxsize=80mm \epsfysize=60mm \epsfbox{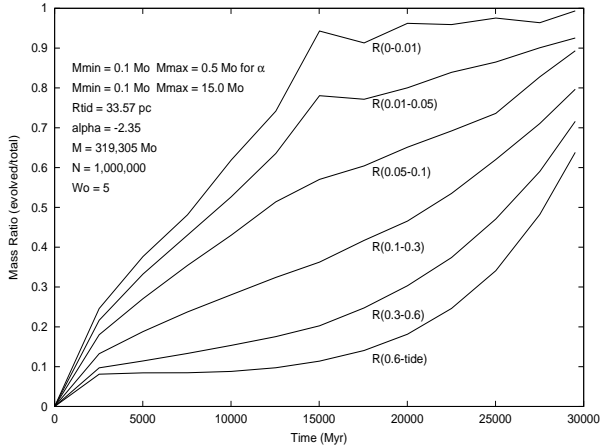}
\end{center}
\caption{Evolution of the ratio of evolved stars to the actual total
mass for different zones: R(0-0.01), R(0.01-0.05), R(005-0.1),
R(0.1-0.3), R(0.3-0.6) and R(0.6-$r_t$) for model W5235.}
\end{figure}
\begin{figure}
\epsfverbosetrue
 \begin{center}
\leavevmode \epsfxsize=80mm \epsfysize=60mm \epsfbox{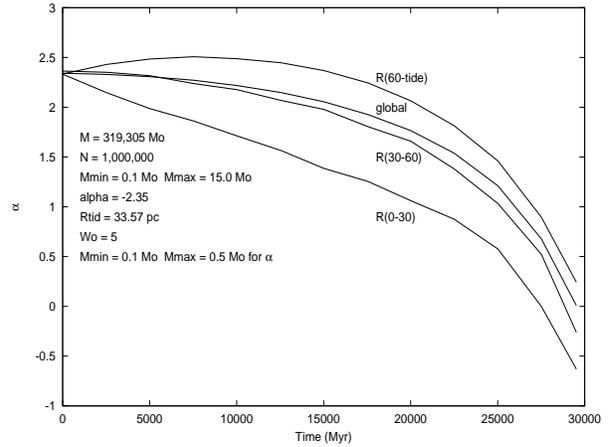}
\end{center}
\caption{Evolution of the power-low index of the mass function for
different shells: R(0-30), R(30-60), R(60,$r_t$) and for the global
mass function for model W5235. The power--low index was computed for
main--sequence stars in the mass range between $0.1 M_\odot$ to $0.5
M_\odot$.}
\end{figure}

Stellar evolution will change a stellar content of the cluster. In
the course of time main--sequence stars will evolve creating first
black holes, neutron stars and then white dwarfs with decreasing
masses. In Figure 9, evolution of the mass ratio of main--sequence
stars ($ms$), white dwarfs ($wd$) and neutron stars/black holes
($ns$) to the actual total cluster mass and the evolution of the
ratio of the actual total mass ($ms+wd+ns$) to the initial total
mass, are shown for model W5235. The fraction of evolved stars
present in the system at any time results from two processes which
act in opposite directions: stellar evolution (formation) and
evaporation across the tidal boundary (loss). At the beginning of
cluster evolution the sharp increase of the mass ratio of evolved
stars and decrease of the mass ratio of main--sequence stars and the
total mass ratio are connected with the stellar evolution of the
most massive stars in the system. Then, the steady increase of the
mass ratio for $wd$ and $ns$ is observed. After the time of core
bounce the ratio for $ns$ is slowly decreasing because of binary
activities (the most massive stars (neutron stars) are with the
highest probability involved in binary formation and finally removed
from the system by interactions with the field stars and other
binaries). The ratios for $ms+wd+ns$ and $ms$ are decreasing, as
expected. At the time close to the dissolution time evolved stars
consists of up to 75\% of the mass of the all stars and up to 63\%
of the number of all stars. The fraction of mass/number of $wd$ is
larger for systems with shorter relaxation time and with stronger
influence of the tidal field of a parent galaxy. This is in a very
good agreement with the results obtained by VH.

Figure 10 shows the evolution of the mass ratio of evolved stars to
the actual total mass for different zones in the system (see
description of Figure 10). As expected, the mass segregation process
and removal of less massive stars due to the evaporation process
across the tidal boundary cause steady increase of the mass ratios.
For the inner part of the system (up to about $5\%$ Lagrangian
radius), during the core collapse time the rate of increase of the
mass ratio is the fastest in the system. However, after the core
bounce the rate becomes slowest The mass ratio is nearly constant.
This is connected with the fact that the mass segregation process is
nearly completed at the time of core bounce (see Figure 7 and its
discussion). Further increase of the mass ratio can be mainly
attributed to binary activities. For parts of the system outside
$10\%$ Lagrangian radius the behaviour of the mass ratio is opposite
to that discussed earlier. During the collapse phase the rate of
mass ratio increase is slower than in the post collapse phase. The
further out $10\%$ Lagrangian radius the ratios are smaller and
faster, respectively. This behaviour can be attributed to the
properties of the stripping process, which is more and more
effective when deeper and deeper parts of the system are exposed. In
between these two zones (Lagrangian radius between $5\%$ and $10\%$)
the rate of increase of mass ratio is nearly constant during the
whole evolution. The amount of mass in evolved stars in the core can
be as high as $97\%$ and in the outer parts of the system up to
$55\%$, at the time close to the dissolution time. The same
qualitative behaviour is observed for other models. However, for
models with steeper initial mass function and for more concentrated
ones the mass ratios are smaller. For initially more concentrated
models the time of core bounce is much shorter than for the less
concentrated models. So, the process of mass segregation is less
advanced and therefore the mass ratios are smaller. Also, for more
concentrated cluster the evaporation process is less effective. The
ratios are about $10\%$ smaller than for W5325 model. For the model
W3235, which does not enter the post--collapse evolution before the
cluster dissolution the mass ratios behave like for other models in
the collapse phase, but the spread between ratios in different zones
is much smaller. It is worth to note that cluster at the time 15 Gyr
can consist of up to $90\%$ of evolved stars in the core and up to
$10\%$ of evolved stars in the outer parts of the system. This
conclusion has very important observational consequences. The
presence of a substantial number of practically invisible stars has
to be taken into account when the global globular cluster parameters
are drown from the observations.

Now lets discuss the evolution of the local (LMF) and global (GMF)
mass functions in comparison to the initial mass function (IMF).
According to VH the mass slope of a mass function can be obtained
from the numerical data in the following way:

\begin{equation}
\alpha = {{ln(dN/dm_1)_1 - ln(dN/dm_2)_2}\over {ln(m_1) - ln(m_2)}}
\end{equation}
$(dN/dm)_{1,2}$ is estimated by $N_{1,2}(t)/\Delta m_{1,2}$, where
$N_{1,2}(t)$ is the total number of stars between $m_{1,2}$ and
$m_{1,2} + \Delta m_{1,2}$. According to VH $m_1$ and $m_2$ were
chosen to $0.1 M_\odot$ and $0.5 M_\odot$, respectively. The LMF
were calculated for three zones (VH): up to $30\%$ Lagrangian
radius, between $30\%$ and $60\%$ Lagrangian radii and from $60\%$
Lagrangian radius up to tidal radius. In the case when only mass
segregation is taken into account, the LMF in the inner parts of the
system becomes flatter and in the outer parts becomes steeper. The
evaporation process acts in the opposite direction. It tends to
flatten the mass function because of the preferential removal from
the system (particularly from the outer parts) of low mass stars. As
it can be seen in Figure 11 the theoretically predicted behaviour of
LMF is qualitatively represented by the numerical results. For all
the discussed models the LMF for the outermost parts of the system
initially increases. The increase is slightly higher for more
concentrated systems with the same IMF (this is consistent with the
picture of mass segregation), and even more pronounced for the
system with the same concentration but with stepper mass function.
For all models the GMF closely resembles the LMF in the middle parts
of the system. The IMF during dynamical cluster evolution is very
quickly forgotten and practically impossible to recover from the
observational data. The actual GMF of globular clusters can be
recovered from observational data for middle parts of the system
(close to the half--mass radius). The results discussed above are in
a very good agreement with the results obtained by VH for $N$--body
runs.

\section{CONCLUSIONS.}

This paper is a continuation of Paper I and II, in which it was
shown that the Monte Carlo method is a robust scheme to study, in an
effective way, the evolution of very large $N$--body systems. The
Monte Carlo method describes in a proper way the graininess of the
gravitational field and the stochasticity of real $N$--body systems.
It provides, in almost as much detail as $N$--body simulations,
information about the movement of any object in the system. In that
respect the Monte Carlo scheme can be regarded as a method which
lies between direct $N$--body and Fokker--Planck models and combines
most advantages of both these methods. This is the first important
step in the direction of simulating the evolution of a real globular
cluster.

It was shown that the results obtained in this paper are in
qualitative agreement with these presented by CW, VH, AH, TPZ, JNR,
Paper II, Baumgardt \& Makino (2002) and Lamers \et 2005.
Particularly good agreement is obtained with VH's $N$--body
simulations and what is not surprising, with results of Monte Carlo
simulations presented in Paper II. All models survive the phase of
rapid mass loss and then undergo core collapse and then subsequent
post--collapse expansion (except model W3235) in a manner similar to
isolated models. The expansion phase is eventually reversed when
tidal limitation becomes important. As in isolated models, mass
segregation substantially slows down by the end of core collapse.
Mass loss connected with stellar evolution dominates the initial
phase of cluster evolution. Then the rate of stellar evolution
substantially slows down and escape due to tidal stripping takes
over. During this phase of evolution the rate of mass loss is nearly
constant, and higher for shallower mass functions. Energy carried
away by stellar evolution events dominates the energy loss due to
tidal stripping, even though the tidal mass loss is higher. For the
first time, because of the large number of particles in simulations
($1,000,000$) which results in substantial reduction of statistical
fluctuations, it was possible to compute in an accurate way the
evolution of density profiles. The development of power--low density
profile is clearly visible and agrees very well with theoretical
predictions. The observed power--low index is equal to about $-2.2$.
The strongly concentrated model (W7235), shows a modest initial
build up of anisotropy in the outer parts of the system. As tidal
stripping exposes the inner parts of the system, the anisotropy
gradually decreases and eventually becomes slightly negative. Model
W5235 from the very beginning develops in the outer parts of the
system negative anisotropy. The cluster is not concentrated enough
to prevent removal of stars which are preferentially on radial
orbits. The negative anisotropy stays negative until the time of
cluster disruption, when it becomes slightly positive (during
cluster disruption most stars are on radial orbits). Because of mass
segregation, and due to evaporation across the tidal boundary, which
preferentially removes low mass stars, the mean mass in the cluster
increases with time. During the core collapse the rate of increase
of the mean mass is highest in the central parts of the system (mass
segregation). After the core bounce there is a substantial increase
in the mean mass in the middle and outer parts of the system (tidal
stripping), and more modest increase in the inner parts of the
system, which is mainly connected with binary activity. The fraction
of evolved stars is increasing during the cluster evolution. This
fraction is larger for systems with shorter relaxation time and
stronger influence of the tidal field of a parent galaxy. The mass
ratio of evolved stars can be as high as nearly $65\%$ at the outer
parts of the system up to nearly $98\%$ in the core. At the time of
$15$ Gyr these ratios are $90\%$ and $10\%$, respectively. The
presence of a substantial number of practically invisible stars has
very important consequences for the interpretation of observational
data and it has to be taken into account when the global globular
cluster parameters are drown from the observations. Because of
stellar evolution, mass segregation and evaporation of stars the IMF
is quickly forgotten and impossible to recover from the
observational data. The actual GMF closely resembles the LMF for the
middle parts of the system (close to half--mass radius). These
results are in an excellent agreement with $N$--body simulations
presented by VH.

In order to perform simulations of real globular clusters the
description of some processes, already included in the code has to
be improved, and several additional physical processes have to be
added to the code. Stellar and binary evolution, more accurate
treatment for energy generation by binaries, particularly in
binary--binary interaction, and proper treatment of the escape
process in the presence of a tidal field are still waiting for
improvement. One of the population synthesis codes developed by
Hurley (Hurley \et 2000, 2002), Portegies Zwart (Portegies Zwart \&
Verbunt 1966) and Belczy\'nski (Belczy\'nski, Kalogera \& Bulik
2002) can be used to follow more accurately single star and binary
evolution. The implementation of techniques used in Aarseth's
$NBODY$ codes for direct integrations of a few body subsystems can
cure the problem of energy generation in 3--body and 4--body
interactions (Giersz \& Spurzem 2003). The treatment of escapers
proposed by Spurzem \et (2005) can be implemented to solve the
problem of escapers in the tidal field (problem which is inherently
connected with gaseous, Fokker--Planck and Monte Carlo codes). The
tidal shock heating of the cluster due to passages through the
Galactic disk, interaction with the bulge, shock--induced
relaxation, primordial binaries, physical collisions between single
stars and binaries are some of the processes, which are waiting for
implementation into the code. The inclusion of all these processes
does not pose a fundamental theoretical challenge, but is rather
complicated from the technical point of view. The international
collaboration called MODEST was setup a few years ago to solve
problems with merging all available codes (hydrodynamical, stelar
evolution and dynamical) into a code which can deal in detail with
the evolution of real globular clusters (see {\bf
http://www.manybody.org/modest}). Inclusion into the Monte Carlo
code of as much as possible physical processes will allow to perform
detailed comparison between simulations and observed properties of
globular clusters, and will also help to understand the conditions
of globular cluster formation and explain how peculiar objects
observed in clusters can be formed. These types of simulations will
also help us to introduce, in a proper way, into future $N$--body
simulations all the necessary processes required to simulate the
evolution of real globular clusters on a star--by--star basis from
their birth to their death.
\bigskip
\bigskip

{\parindent=0pt {\bf Acknowledgments} I would like to thank Douglas
C. Heggie and Rainer Spurzem for stimulating discussions, comments
and suggestions. This work was partly supported by the Polish
National Committee for Scientific Research under grant 1 P03D 002 27.}

\end{document}